\documentclass[conference]{IEEEtran}

\usepackage{cite, makecell}
\usepackage{amsmath,amssymb,amsfonts}
\usepackage{algorithmic}
\usepackage{graphicx}
\graphicspath{{images/}}
\usepackage{textcomp}
\usepackage{xcolor, rotating}
\usepackage{balance}
\usepackage{multirow}
\usepackage{gensymb}
\usepackage{array,booktabs,longtable,tabularx,tabulary}
\usepackage{lipsum}
\makeatletter
\newcommand\singlelipsum[1]{%
  \begingroup\let\lips@par\relax\csname lipsum@\@roman{#1}\endcsname
\endgroup }
\makeatother

\def\BibTeX{{\rm B\kern-.05em{\sc i\kern-.025em b}\kern-.08em
		T\kern-.1667em\lower.7ex\hbox{E}\kern-.125emX}}

\usepackage{amssymb,amsmath,epsfig,latexsym, bm}
\usepackage{nohyperref}	
\hypersetup{colorlinks, pdfa=true, breaklinks, citecolor=black, urlcolor=black, linkcolor=black}
\usepackage{amssymb,amsmath,epsfig,latexsym, bm}
\usepackage{hyphenat}

\usepackage{mathtools, mdwmath, mdwtab, multirow, amssymb, amsmath, array, ragged2e}

\newcommand*{\email}[1]{\normalsize\texttt{\href{mailto:#1}{#1}}\par}

\usepackage{threeparttable, booktabs, url, eqparbox, xspace, setspace, tabularx, subfig}
\makeatletter
\DeclareRobustCommand\onedot{\futurelet\@let@token\@onedot}
\def\@onedot{\ifx\@let@token.\else.\null\fi\xspace}

\def\etc{\emph{etc}\onedot} 
 
\def\etal{\emph{et al}\onedot}
\makeatother

 \usepackage[pscoord]{eso-pic}
\newcommand{\placetextbox}[3]{
 \setbox0=\hbox{#3}
 \AddToShipoutPictureFG*{ \put(\LenToUnit{#1\paperwidth},\LenToUnit{#2\paperheight}){\vtop{{\null}\makebox[0pt][c]{#3}}}
 }
 }

 \placetextbox{.21}{0.055}{\small{978-1-6654-6658-5/22/\$31.00 ©2022 IEEE}}

\begin{document}

\title{A Transfer-Learning Based Ensemble  Architecture for ECG Signal Classification }


\author{
    \IEEEauthorblockN{%
         Tareque Bashar Ovi\IEEEauthorrefmark{1},
         Sauda Suara Naba\IEEEauthorrefmark{2},
         Dibaloke Chanda\IEEEauthorrefmark{3} and
         Md. Saif Hassan Onim \IEEEauthorrefmark{4}
                    }

    \IEEEauthorblockA{%
        Department of Electrical, Electronic and Communication Engineering,\\
        Military Institute of Science and Technology (MIST) Dhaka-1216, Bangladesh}
    
    \IEEEauthorrefmark{1}\email{ovitareque@gmail.com},
    \IEEEauthorrefmark{2}\email{suaranaba@gmail.com},
    \IEEEauthorrefmark{3}\email{dibaloke66@gmail.com}
    \IEEEauthorrefmark{4} \email{onim.hassan16@gmail.com}
                
    }
\maketitle

\begin{abstract}

Manual interpretation and classification of ECG signals lack both accuracy and reliability. These continuous time-series signals are more effective when represented as an image for CNN-based classification. A continuous Wavelet transform filter is used here to get corresponding images. In achieving the best result generic CNN architectures lack sufficient accuracy and also have a higher run–time. To address this issue, we propose an ensemble method of transfer learning-based models to classify ECG signals. In our work, two modified VGG-16 models and one InceptionResNetV2 model with added feature extracting layers and ImageNet weights are working as the backbone. After ensemble, we report an increase of $6.36\%$ accuracy than previous MLP based algorithms. After 5-fold cross-validation with the Physionet dataset, our model reaches an accuracy of $\mathbf{99.98\%}$.

\end{abstract}
\color{black}

\section{Introduction}

\begin{figure}[h]
    \centering
    \includegraphics[width=.8\columnwidth]{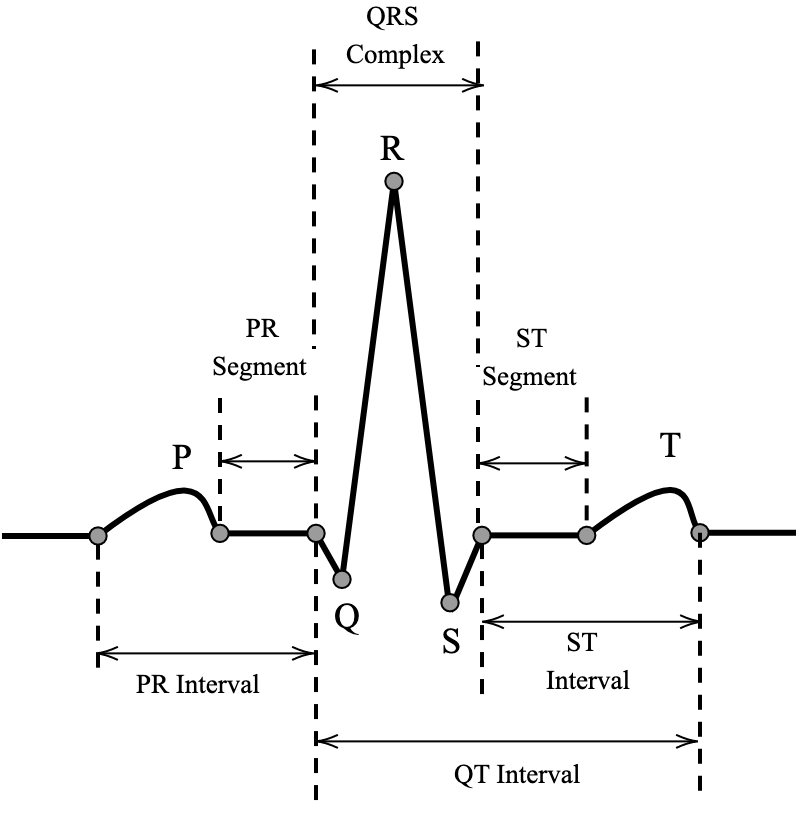}
    \caption{Typical ECG Signal}
    \label{ecg_signal}
\end{figure}

ECG documents the electrical function of the heart by encapsulating essential electrical activities. It is crucial for the preliminary detection of numerous cardiac abnormalities and cardiovascular diseases.
Substantial facts like a patient’s heart rate, heart rhythm, record of heart attack, probable narrowing of the coronary arteries, traces of reduced oxygen distribution to the heart, and so forth can be attained from an ECG report. 
ECGs are usually conducted in three basic ways: \textit{resting ECG, ambulatory ECG}, and \textit{cardiac stress test}. An ECG graph (Fig \ref{ecg_signal}) can be characterized by a repeated series of \textit{P, QRS, T,} and a conditional \textit{U} wave. The peaks (\textit{P, Q, R, S, T} and \textit{U}), intervals (\textit{PR, RR, QRS, ST}, and\textit{ QT}), and segments (\textit{PR} and \textit{ST}) represent the traits of an ECG.
Deviation from the expected form of electrical signals signifies several issues including defects or abnormalities in the \textit{heart’s shape, size, electrolytes, ischemia, heart rhythm} \etc. 
Once the anomaly is detected, risk factors for cardiovascular diseases can be diagnosed and treated, consequently upgrading the patient's medication. 

However, unlike detecting, classifying the abnormalities in ECG precisely is an immensely complicated task considering the numerous challenges that arise throughout the process.
One of the challenges can be examining and identifying cardiac problems from ECG paper. Again, applying ordinary visual analysis approaches can be challenging for physicians as well.
Resultantly, intelligent computer systems that utilize sophisticated image and signal processing techniques have been introduced. This aids in more precise diagnosis of patients and offers better treatment than empirical procedures.
Even in the case of a fully automated and computerized system, it's not an easy process, considering the presence of external noise and uneven classes in the dataset.
The accuracy achieved applying machine learning is insufficient as well. Multiple attempts have been made to incorporate deep learning algorithms to address and solve the issue of precisely classifying cardiovascular diseases with a lower number of false positive detections.

Zihlmann~\etal~\cite{zihlmann2017convolutional} presented two neural network architectures, a convolutional neural network (CNN) and a convolutional recurrent neural network (CRNN), for ECG classification. On the hidden challenge testing set, the second architecture exceeded the first in terms of accuracy, with an F1 score of 82.1 percent for the PhysioNet/CinC Challenge 2017 atrial fibrillation (AF) classification data set. Rajkumar~\etal~\cite{rajkumar2019arrhythmia} employed CNN to attain the features automatically from the time domain ECG signals collected from the MIT-BIH database~\cite{goldberger2000bih}. They achieved 93.6 percent accuracy using the ELU activation function. Syama~\etal~\cite{syama2019classification} implemented a multilayer-perceptron neural network (MLPNN) for the classification of ECG signals collected from the MIT-BIH database and achieved 94 percent accuracy. ESSA~\etal~\cite{essa2021ensemble} came up with a deep learning-based multi-model ensemble by implementing a combination of a convolutional neural network (CNN) and long short-term memory (LSTM) network in order to classify ECG. The proposed approach was evaluated on the MIT-BIH database to derive the experimental findings with an overall accuracy of 95.81 percent. Boussaa~\etal~\cite{boussaa2016ecg} employed Mel Frequency Coefficient Cepstrum algorithm (MFCC) and artificial neural network (ANN) classifier for classifying ECG signal extracted from the MIT-BIH database. Hannun~\etal~\cite{hannun2019cardiologist} collected 91,232 single-lead ECGs from 53,549 individuals who used a single-lead ambulatory ECG monitoring device to create a deep neural network (DNN) to distinguish 12 rhythm categories. ŞEN~\etal~\cite{csen2019ecg} proposed ECG time-series signal classification and ECG spectrogram pictures to classify heartbeat arrhythmia.
Ullah~\etal~\cite{ullah2020classification} proposed a two-dimensional (2-D) convolutional neural network (CNN) model in order to classify ECG signals into eight classes and achieved an accuracy of 99.11 percent. Cordoș~\etal~\cite{cordo2021} proposed a strategy to use ECG signal for biometric identification employing Inception-v3, Xception, MobileNet and NasNetLarge models. They found Inception-v3 model to be the most effective model for ECG classification having an accuracy of 99.5 percent.
Gajendran~\etal~\cite{gajendran2021ecg} used different modern deep networks trained on the ImageNet database to categorize scalograms (2D representations) of ECG signals collected from three PhysioNet databases. Rahuja~\etal~\cite{rahuja2021deep} introduced a classifier for automatic ECG classification employing continuous wavelet transform and AlexNet CNN. They used PhysioNet research dataset for their experiment and attained an overall accuracy of 97.3 percent with a limited dataset. Their works demonstrate that modern CNN architectures performs well for multi-class ECG signal classification \cite{demonbreun2020automated, zhang2018ecg, ahamed2020ecg, huang2020accurate}.

Our major contribution is introducing modified VGG-16 and InceptionResNetV2 models pretrained with ImageNet weights and finally constructing a soft voting based ensemble architectures for taking maximum accuracy for respective classes.

\section{Methodology}

\subsection{Dataset Description}
\begin{figure}[h]
    \centering
    \includegraphics[width=\columnwidth]{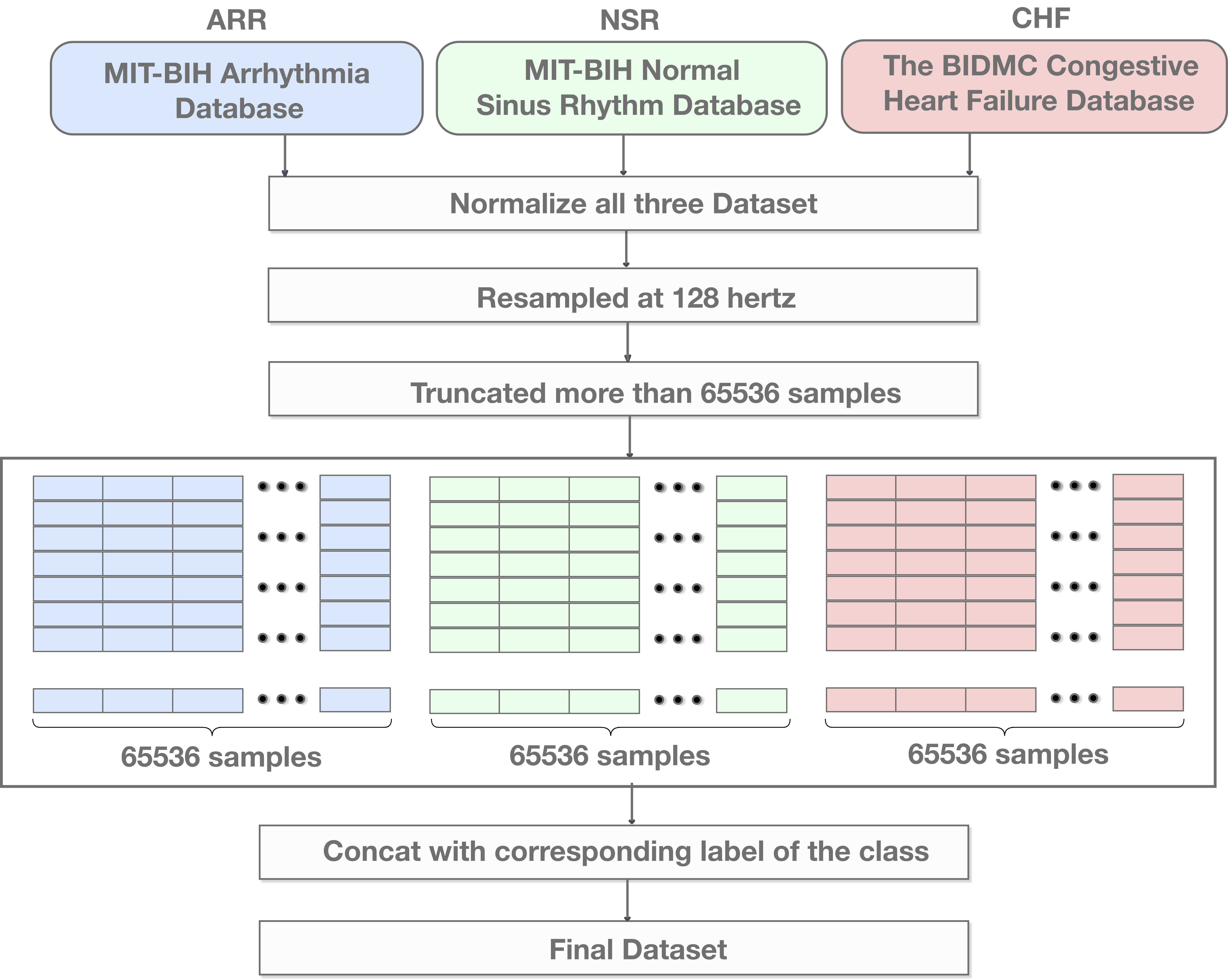}
    \caption{Modified Dataset Combining MIT-BIH Arrhythmia Database, MIT-BIH Normal Sinus Rhythm Database, BIDMC Congestive Heart Failure Database}
    \label{Modified Dataset}
\end{figure}

We used a pre-existing open-source dataset for this work which consists of three distinct types of ECG signals from three different types of heart condition \textit{Arrhythmia (ARR)}, \textit{Normal Sinus Rhythm (NSR)}, and \textit{Congestive Heart Failure (CHF)}. The original version of the dataset is available at PhysioNet\cite{goldberger2000physiobank} in three separate sections  which are MIT-BIH Arrhythmia Database \cite{moody2001impact}, MIT-BIH Normal Sinus Rhythm Database\cite{goldberger2000bih} and BIDMC Congestive Heart Failure Database \cite{baim1986survival}. The modified version of the dataset  combines these three  separate databases  in a common format as shown in fig \ref{Modified Dataset}.

A common scaling factor was used to normalize the range of the magnitude followed by a sampling stage with a common sampling rate of $128Hz$ and number of samples equal to  $65536$. This allowed all signals to have the same length. Number of instances for ARR, NSR, and CHF class are respectively $96$, $30$, and $36$ which equates to a total of $162$ ECG recordings.

 \begin{figure*}[t]
    \centering
    \includegraphics[width=1.6\columnwidth, height=2in]{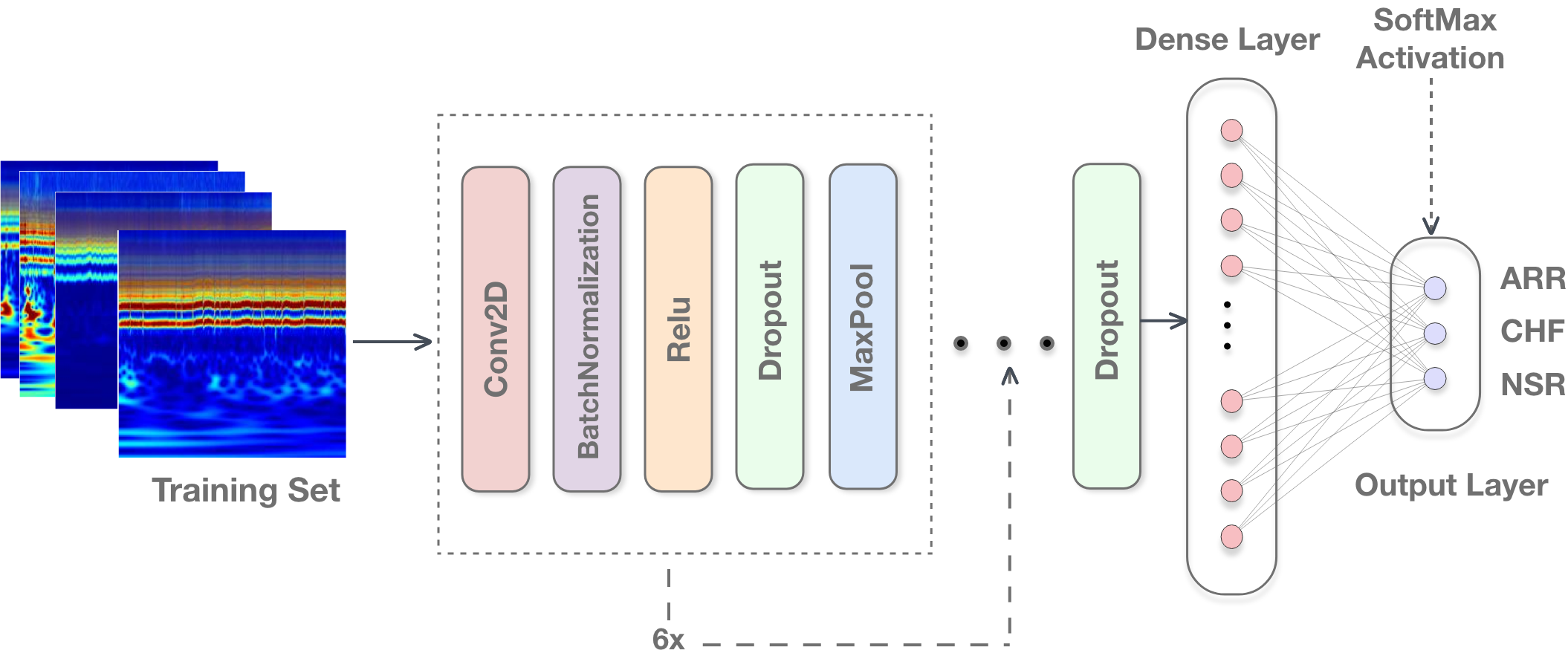}
    \caption{Custom Convolutional Neural Network Architecture}
    \label{simple_cnn}
\end{figure*}

\subsection{Dataset Preprocessing}

Time series data is not compatible  with CNN architecture. To make it compatible, CWT(Continuous Wavelet Transform) is utilized to convert the one-dimensional signal to a two-dimensional scalogram.

\begin{equation}
\label{cwt}
C(a, b ; f(t), \psi(t))=\int_{-\infty}^{\infty} f(t) \frac{1}{a} \psi^{*}\left(\frac{t-b}{a}\right) d t
\end{equation}

\begin{figure}
    \centering
    \includegraphics[width=\columnwidth, height=2.8in]{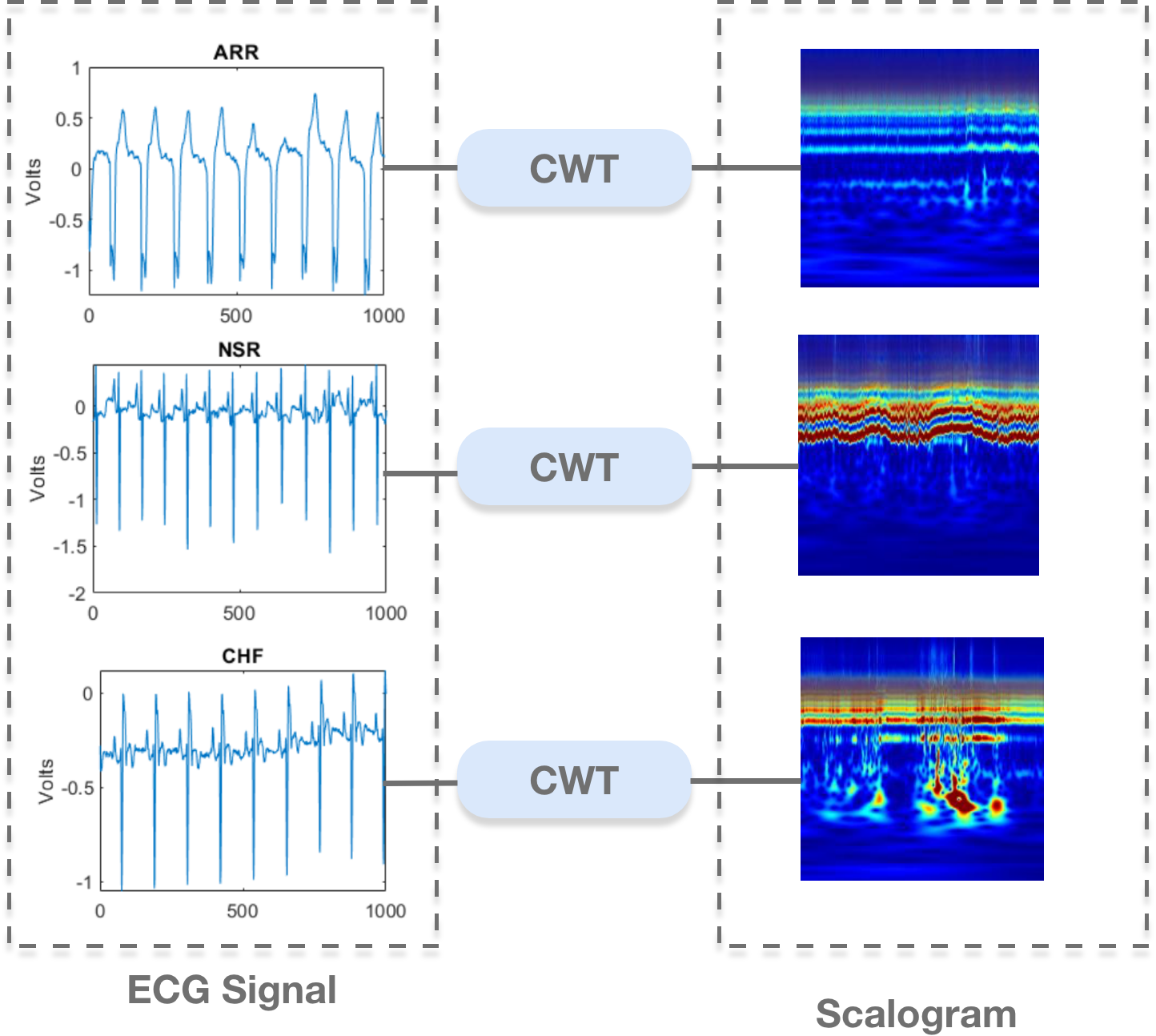}
    \caption{Conversion from ECG Signal to Scalogram  Using Continuous Wavelet Transform}
    \label{signal_to_image}
\end{figure}

CWT is generally expressed by the equation shown in \eqref{cwt} defined by  a scale parameter $a>0$ and position parameter $b$. This is used as an analyzing function to measure the  similarity between the wavelet and the signal by equating the inner product between them. The CWT is defined as the sum of the signal multiplied by scaled and shifted versions of the wavelet function $\psi(t)$ also known as the mother wavelet. A small value of $a$ signifies a compressed wavelet which captures rapid changing details whereas a large value of $a$ captures slowly changing details. Generated scalograms after CWT applied to 1D signals can be interpreted as a time-frequency representation of the time-series signal. Fig \ref{signal_to_image} shows sample scalogram images after applying CWT to three ECG classes. The generated scalograms are of dimension of $224 \times 224 \times 3$ each.

\subsection{Proposed Methods}
 
 Two different approaches are explored to correctly categorize the three classes. Firstly, a simple CNN  architecture is used as a reference to classify the scalograms. In the second approach, we made use of both transfer learning and ensemble methods combining three different pre-trained deep CNN architectures. The following sections will give a detailed overview of these two approaches.

 For the first approach, a CNN is constructed by stacking several repeated blocks as shown in fig \ref{simple_cnn}. A single block consists of a \textit{convolutional layer} followed by a \textit{Batch Normalization} layer. The purpose of the \textit{convolutional layer} is to extract relevant features from the scalogram. \textit{Batch normalization layers} significantly speed up the training process as it standardizes the input data for each batch. The next layer is a \textit{dropout layer} whose purpose is to prevent overfitting by randomly turning off some learning parameter. The final layer in the block is a \textit{pooling layer} which reduces the spatial size of the previous layer output. These repeated blocks are finally terminated with an additional \textit{dropout layer}. After that, flattening is done to convert to \textit{dense layer} which is required for classification. The output layer is activated by \textit{Softmax Activation Function}. Hyper-parameters used for this model is summarized in Table \ref{parameters table}. They were fine tuned after several test and trial to keep high accuracy and low loss.

\begin{table}[h]
\caption{Training Parameters for  Convolutional Neural Network}

\centering
\renewcommand{\arraystretch}{1}
	\setlength{\tabcolsep}{6pt}
	\resizebox{\columnwidth}{!}{

    \begin{tabular}{lc}
    \toprule
    \textbf{Parameter} & \textbf{Value}\\
    \midrule
    Base Architecture & CNN 
    \\
    Classes & 3 \\ 
    Input Dimension & $224 \times 224 \times 3$ \\
    Number of Epochs Trained & 200 \\
    Model Type & Multiclass Classification \\
    Hidden Layer Activation & Rectified Linear Unit (ReLU) \\ 
    Output Layer Activation & Softmax \\
    Optimizer & Adam \\
    Initial Learning Rate & 0.05\\ 
    Dropout Rate  & 0.5 \\
    Padding Type & Same \\
    Shuffle Frequency & Each Epoch \\ 
   
    \bottomrule
    \end{tabular}
    }
\label{parameters table}

\end{table}

\begin{table}[!ht]
\small
\caption{Augmentation parameters for augmentation of training images }
\centering
\renewcommand{\arraystretch}{1}
	\setlength{\tabcolsep}{15pt}
	\resizebox{\columnwidth}{!}{
    \begin{tabular}{lccc}
        \toprule
        \textbf{Augmentation Parameter} &  & \textbf{Value}\\
        \midrule
        Random Flip & - & $Horizontal$ \\
        Random Rotation & - & $0.2$ \\
        \bottomrule
    \end{tabular}
    }
\label{augmentation table}
\end{table}

\begin{figure*}[t]
    \centering
    \includegraphics[width=1.8\columnwidth, height=3in]{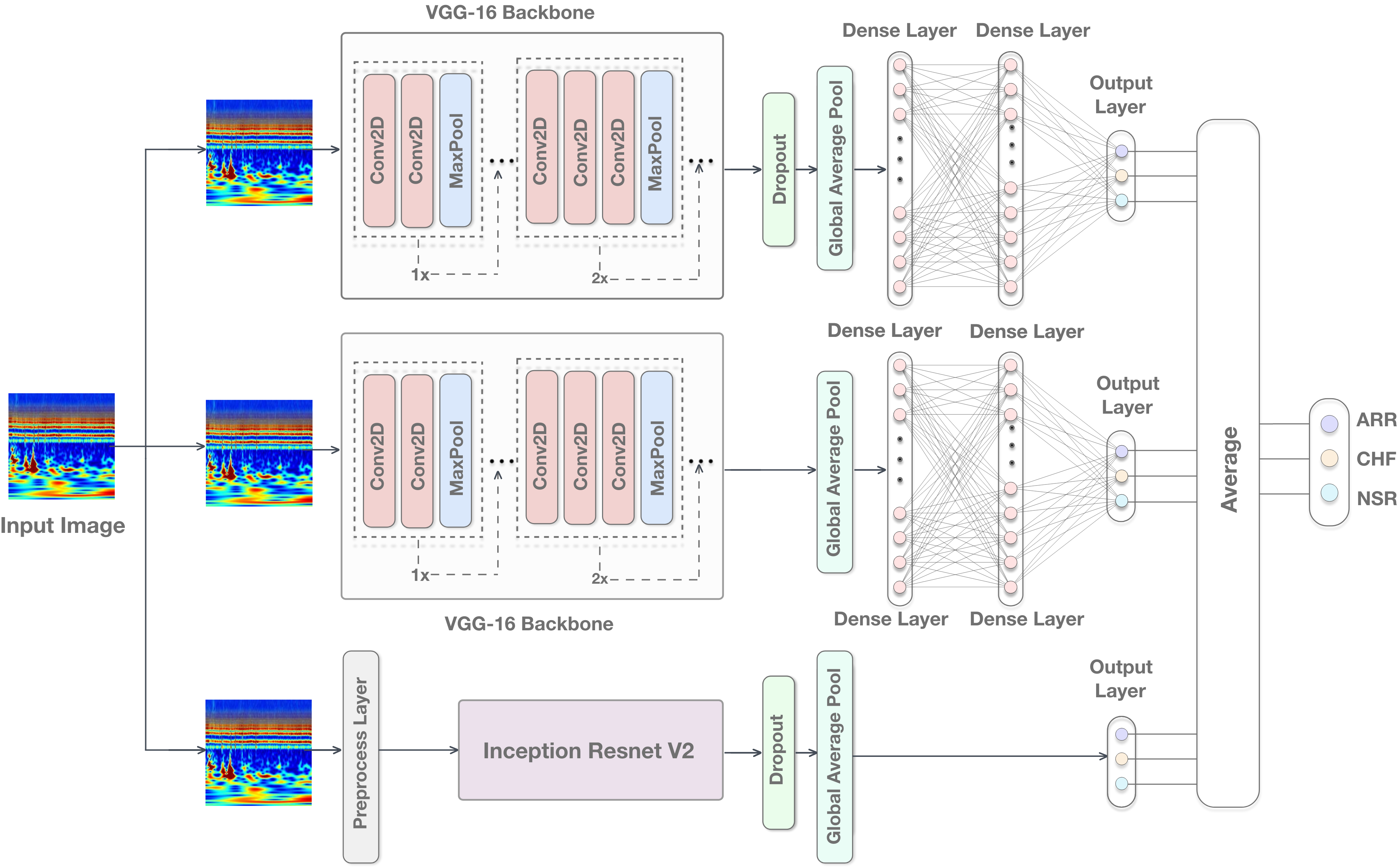}
    \caption{Architecture of the ensemble model}
    \label{model_arch}
\end{figure*}

The final layer is composed of three perceptions and activation function for this layer is softmax, which is a common practice for multi-class classification. Softmax generates a  probability value between 0 to 1 based on the confidence of prediction.The model was trained for 4 min 36 sec with a batch size of 128.

In the second approach, prior to building the model, augmentation was performed to increase the quantity of training data and to incorporate variance into the dataset.
The dataset's size expanded from 162 to 486 as a result of the augmentation. 

Two different state-of-the-art  pre-trained CNN architectures are incorporated for the ensemble model which are, respectively, VGG-16 and InceptionResNetV2.  A brief description about these two models are given in the following section.

\subsubsection{Modified VGG-16} 
VGG-16 is a CNN Architecture that won the 2014 ImageNet Large Scale Visual Recognition Challenge (ILSVRC). The VGG-16 architecture is an enhancement to the AlexNet design. Instead of directly employing the VGG-16, we made some slight modification, but kept the feature extraction part of the model which consists of consecutive convolutional and pooling layers. Instead of directly transitioning to fully connected layers, two intermediate layers, dropout and  global average pooling are used. Fully connected layers are vulnerable to overfitting, limiting the entire network's generalization capabilities. To avoid this, dropout is added as an intermediate layer for regularization. The reason behind using the global average pooling layer is it enforces a strong correspondence between the feature maps and categories. Finally, three fully connected dense layers are used, where the final layer is comprised of three perceptions for three categories.

\subsubsection{Modified InceptionResnetV2} InceptionResNetV2 is a convolutional neural network that extends the Inception family but includes residual connections. It was trained on millions of images from the \textit{ImageNet} Database. The architecture accepts data in a certain format. Because of this, a preprocessing layer is used to convert to that particular format. Dropout and global average pooling are used for the same reason mentioned above. Finally a single dense layer with softmax activation is used for classification.The complete ensemble architecture is shown in fig \ref{model_arch}.

Ensemble method combines multiple architectures or algorithms to make predictions by combining the outputs. For classification type problem there are two major ways to combine the outputs, one is voting and another is averaging. In our work, we implemented the latter approach. The averaging mechanism works by combining each model's prediction equally.
The hyperparameters for the ensemble model are summarized in Table \ref{ensemble_parameters table}.

\begin{table}[h]
\caption{Training Parameters for Ensemble Architecture}

\centering
\renewcommand{\arraystretch}{1}
	\setlength{\tabcolsep}{10pt}
	\resizebox{\columnwidth}{!}{

    \begin{tabular}{lc}
    \toprule
    \textbf{Parameter} & \textbf{Value}\\
    \midrule
  
    Number of Epochs Trained & 80 \\
    Batch Size & 1 \\
    Output Layer Activation & Softmax \\
    Optimizer & Adam \\
    Initial Learning Rate & 0.05\\ 
    Loss Function & Categorical Crossentropy\\
    Total Number of Parameters & 91,391,817 \\
    Trainable Parameters & 37,020,649 \\
    Non-Trainable Parameters & 54,371,168 \\
   
    \bottomrule
    \end{tabular}
    }
\label{ensemble_parameters table}

\end{table}

\section{Result Analysis}

With the custom CNN approach, the accuracy value is not satisfactory enough. The reason behind such performance is mainly the dataset size. Such a small dataset is not suitable for machine learning models because they tend to overfit, meaning they perform really well on training data, but poorly on test data. Even though the training accuracy is nearly $100\%$, the validation accuracy is  $59.38\%$. We tried several empirical methods to regularize the model, but the accuracy value did not improve.

Because of this, in the second approach we utilized data augmentation. Additionally, after analyzing prior studies, we came to the conclusion that transfer learning works well for scalograms and can achieve quite good accuracy  which motivated us to incorporate transfer learning in ensemble methods.

\begin{figure}[tbp]  
    \centering
  
    \subfloat[Accuracy vs Epoch\label{fig:ex1orig}]{%
       \includegraphics[width=0.48\linewidth]{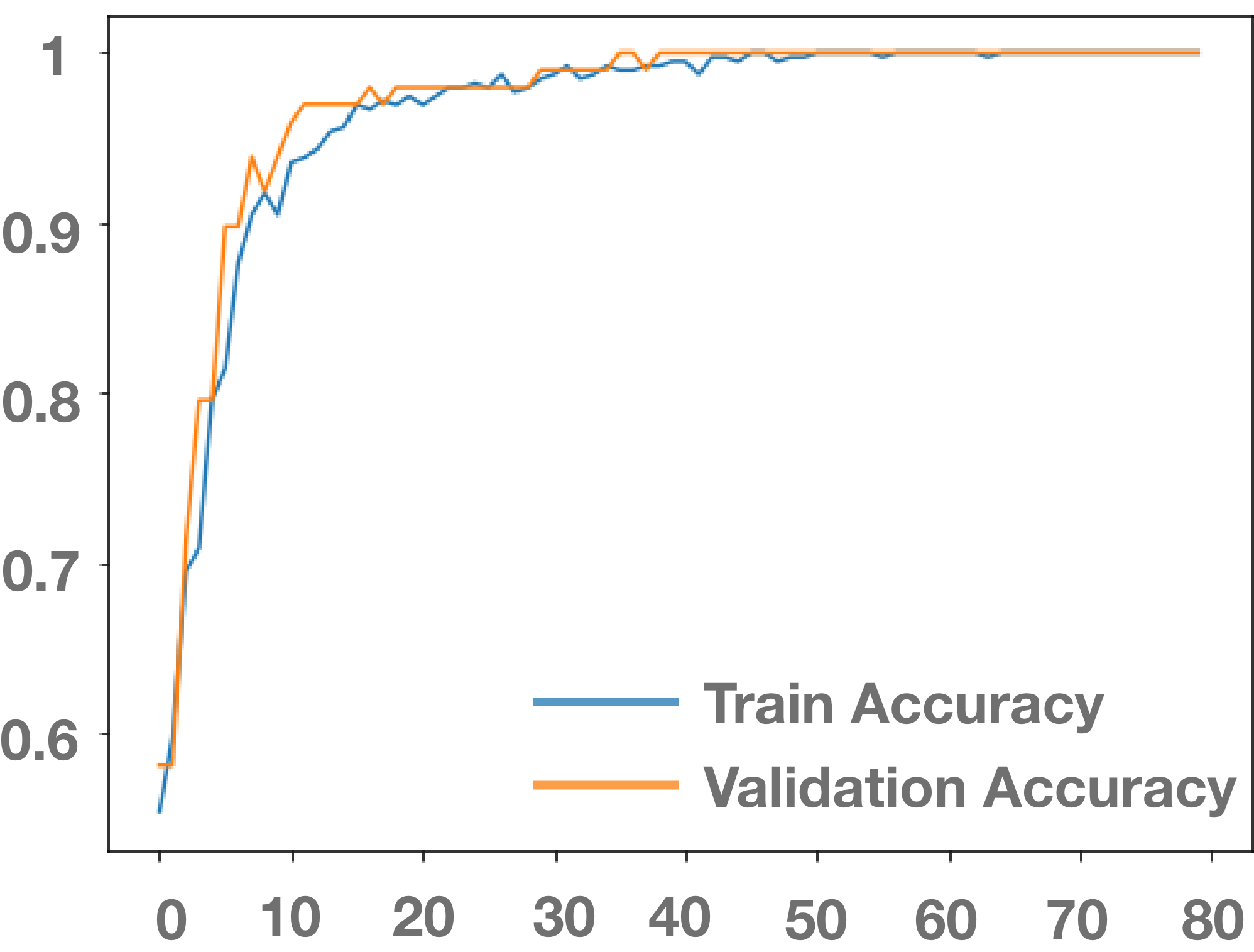}}
    \hfill
    \subfloat[Loss vs Epoch\label{fig:ex1gt}]{%
        \includegraphics[width=0.48\linewidth]{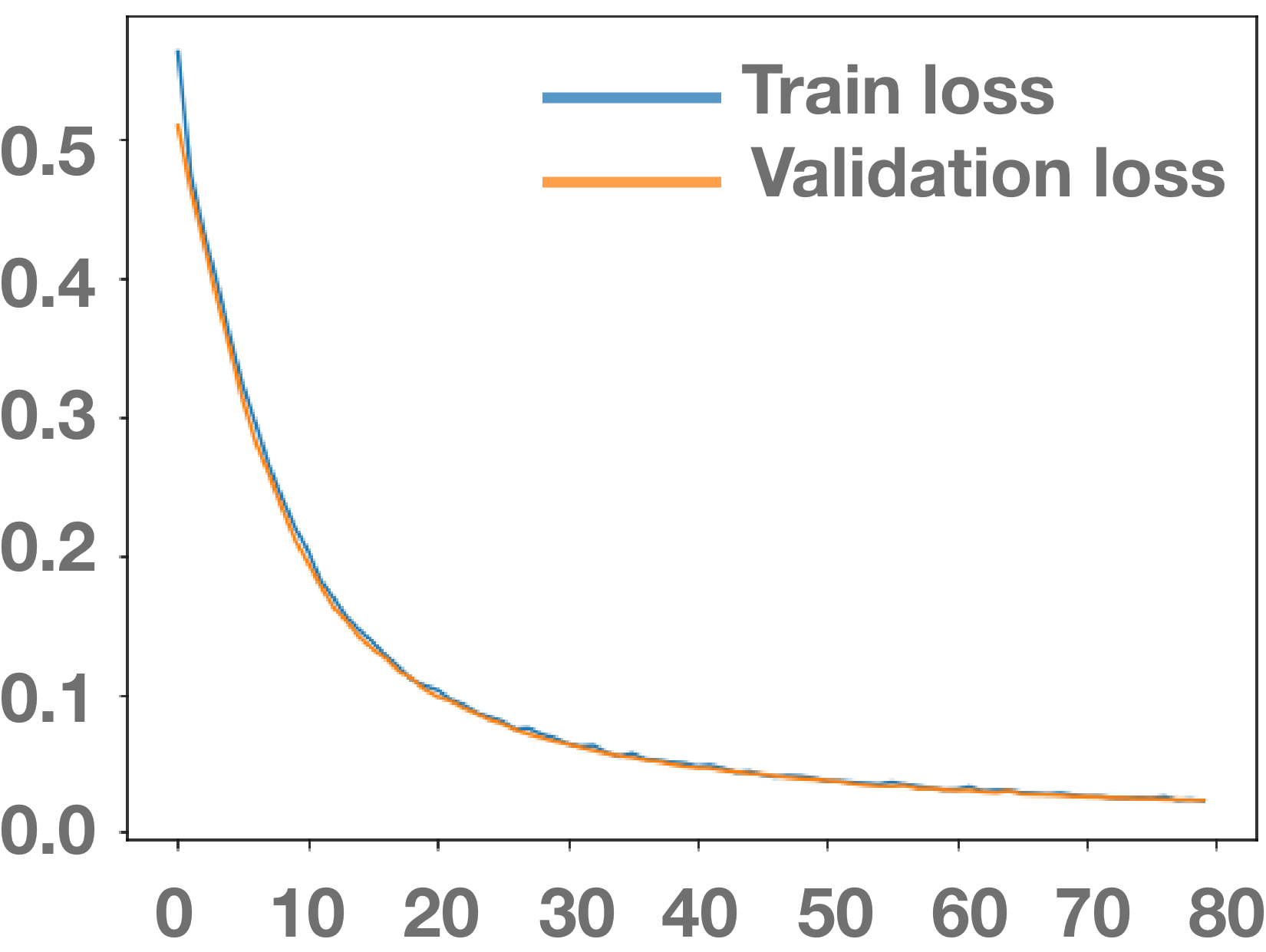}}
    \hfill
   
    \caption{Training and validation Evaluation Curves for Ensemble Architecture}
    \label{Training and validation Evaluation Curves for Ensemble Architecture} 
\end{figure}

\begin{equation}
\label{prec}
\text { Precision }=\frac{T P}{T P+F P}
\end{equation}

\begin{equation}
\label{rec}
\text { Recall }=\frac{T P}{T P+F N}
\end{equation}

Fig \ref{Training and validation Evaluation Curves for Ensemble Architecture} represents how the accuracy and loss value change with increase in epoch for the ensemble approach. From the curves, it is obvious that the validation performance of the model is almost as good as the training performance which means the model is able to generalize. To make sure the model did not overfit, we performed k-fold cross validation for $20$ epoch  with $k=5$ which resulted in around $99.98\%$ accuracy.

\begin{figure}[h]
    \centering
    \includegraphics[width=\columnwidth]{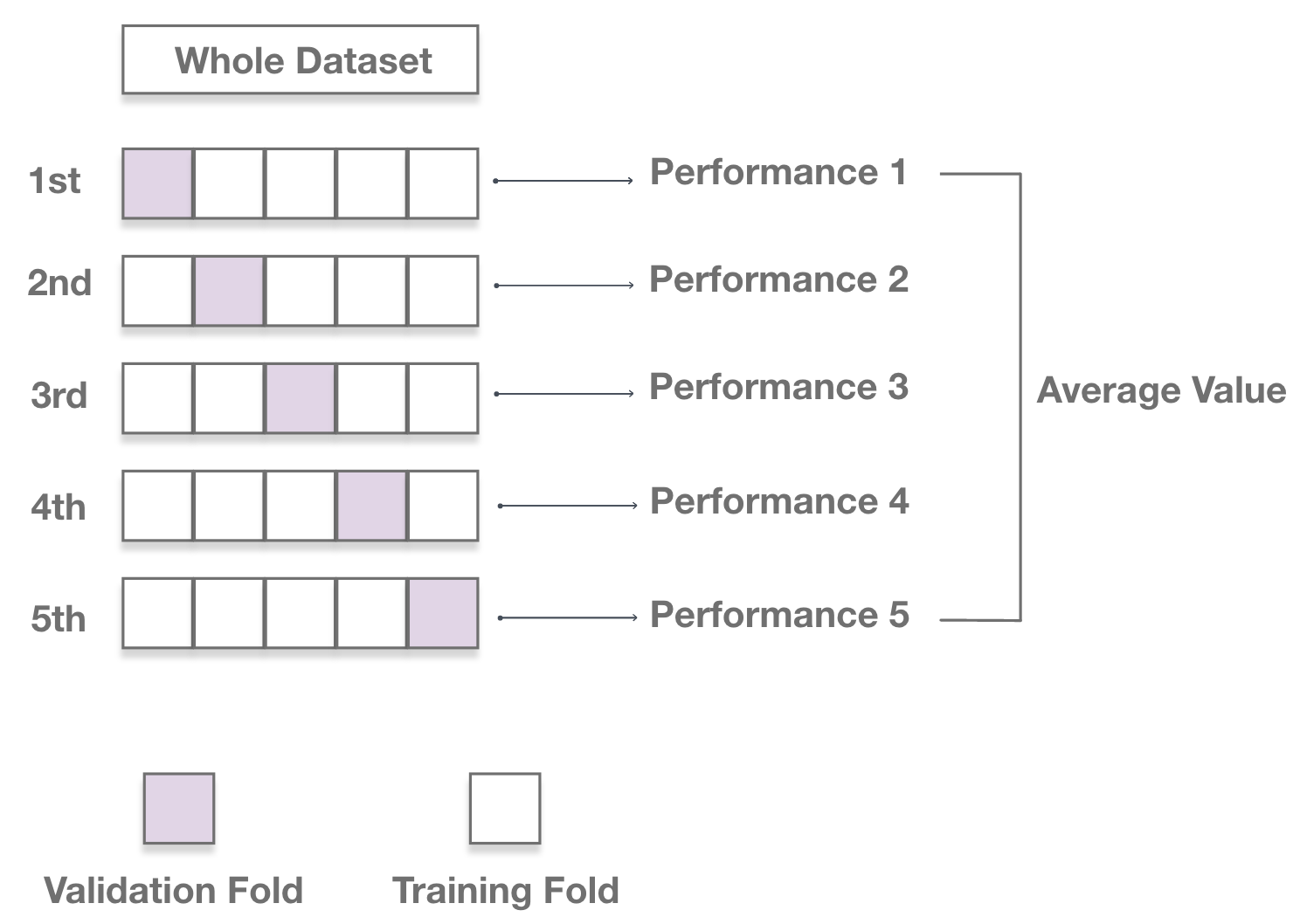}
    \caption{Illustration of 5-fold cross validation}
    \label{k_fold}
\end{figure}

 Along with accuracy, other evaluation metrics like loss, accuracy, precision, and recall are calculated.

 \begin{table}[!ht]
\small
\caption{Evaluation Metrics of the Validation data for the Ensemble model}
\centering
\renewcommand{\arraystretch}{1}
	\setlength{\tabcolsep}{22pt}
	\resizebox{\columnwidth}{!}{
    \begin{tabular}{lccc}
        \toprule
        \textbf{Evaluation Metric} &  & \textbf{Value}\\
        \midrule
        Loss & - & $0.226$ \\
        Accuracy & - & $1.0$ \\
        
        Precision & - & $1.0$ \\
        
        Recall & - & $1.0$ \\
        \bottomrule
    \end{tabular}
    }
\label{eval}
\end{table}

 \textit{ Precision} and \textit{recall} are defined by the ratio shown in equation \eqref{prec} and \eqref{rec} respectively. Where \textit{TP = True Positive, FP = False Positive} and \textit{FN = False Negative.}
 All the evaluation metrics are summarized in Table \ref{eval}.

 Table \ref{tab:comp} compares the performance of some state-of-the-art
research works  with our work.

\begin{table*}[htbp]
	\centering
	\renewcommand{\arraystretch}{1.3}
	\setlength{\tabcolsep}{20pt}
	\caption{Performance Comparison with Existing Work \label{tab:comp}}
	\resizebox{.8\textwidth}{!}{
		\begin{tabular}{lllll}
			\toprule
			\textbf{Year} & \textbf{Reference} & \textbf{Dataset} &  \textbf{Method}  & \makecell[l]{\textbf{Performance}\\\textbf{(Accuracy)}} \\
			
			\midrule
				2018 & Zhang \etal \cite{zhang2018ecg} & MIT-BIH & 12--layer 1--d CNN & $97.70\%$ \\

			\midrule
    			2019 & Syama \etal \cite{syama2019classification} & MIT-BIH & MLPNN &  $94.00\%$ \\

			\midrule
				2020 & Huang \etal \cite{huang2020accurate} & MIT-BIH  & \makecell[l]{FCResNet\\ using MOWPT} & $98.79\%$  \\

    		\midrule
    			2020 & Ahamed \etal \cite{ahamed2020ecg} & MIT-BIH & \makecell[l]{KNN \\ Decision Tree \\ 
                ANN  \\ Support Vector Machine \\ 
                LSTM \\  Ensemble Approach}   & \makecell[l]{$97.64\%$ \\ $96.11\%$ \\  $98.06\%$ \\ $97.58\%$ \\ $95.53\%$ \\ $97.78\%$ } \\

			\midrule
				2021 & Cordoș \etal \cite{cordo2021} & Fantasia  & \makecell[l]{Transfer Learning\\ with Inception-v3} & $99.5\%$ \\

			\midrule
            	\textbf{2022} & \textbf{Ours}  &  \textbf{Physionet}  &  \makecell[l]{\textbf{Transfer Learning based}\\ \textbf{Ensemble with}\\ \textbf{Data Augmentation}} & $\mathbf{99.98\%}$ \\
			
			\bottomrule
			
		\end{tabular}
 	}
\end{table*}



  



\topskip 10pt

\section{Conclusion}
Our model achieves $99.98\%$ accuracy with 5-fold cross validation to classify three classes ARR, NSR and CHF experimented on Physionet dataset. Time series ECG data was transformed into image data with wavelet transformation for compatibility with CNN models. CNN based architectures generated comprehensive results whereas transfer learning-based model performed better. But single transfer learning-based model was not adequate enough to command perfect accuracy and reliability. Hence, our proposed transfer learning based ensemble model with soft voting outperforms all previous literature with a cost of minor computational run-time.

\bibliographystyle{./bibliography/style}
\bibliography{./bibliography/reference}

\end{document}